# Multi-Revolution Low-Thrust Trajectory Optimization Using Symplectic Methods

Zhibo E[1], Davide Guzzetti[2]

*Tsinghua University, Beijing, China, 100084*

**Abstract:** Optimization of low-thrust trajectories that involve a larger number of orbit revolutions is considered a challenging problem. This paper describes a high-precision symplectic method and optimization techniques to solve the minimum-energy low-thrust multi-revolution orbit transfer problem. First, the optimal orbit transfer problem is posed as a constrained nonlinear optimal control problem. Then, the constrained nonlinear optimal control problem is converted into an equivalent linear quadratic form near a reference solution. The reference solution is updated iteratively by solving a sequence of linear-quadratic optimal control sub-problems, until convergence. Each sub-problem is solved via a symplectic method in discrete form. To facilitate the convergence of the algorithm, the spacecraft dynamics are expressed via modified equinoctial elements. Interpolating the non-singular equinoctial orbital elements and the spacecraft mass between the initial point and end point is proven beneficial to accelerate the convergence process. Numerical examples reveal that the proposed method displays high accuracy and efficiency.

---

[1] Corresponding author, School of Aerospace Engineering, ezb16@mails.tsinghua.edu.cn.
[2] School of Aerospace Engineering, dguzzett@mail.tsinghua.edu.cn.

# 1 Introduction

Low-thrust electric propulsion systems have attracted a significant amount of research interest in recent years, owing to a specific impulse higher than traditional chemical propulsion. Thus, low-thrust electric propulsion systems typically consume less fuel mass and, as a result, they are an important option for interplanetary missions. Successful utilization of low-thrust electric propulsion in interplanetary missions includes Deep Space 1 (Rayman et al. 1999), Dawn (Rayman et al. 2007), Hayabusa (Kuninaka et al. 2005), etc. Unfortunately, the application of low thrust usually results in long-duration orbit transfers, which may involve hundreds or even thousands of orbit revolutions. Due to such characteristic geometry, optimizing low-thrust multi-revolution transfers has been considered a challenging problem since several decades ago, and the identification of high-performance transfer optimization frameworks is still an ongoing process. To alleviate the computational effort, developing high-precision and efficient algorithms to optimize transfers with a large number of orbit revolutions is considered to be of great significance.

Numerous computational methods for solving low-thrust optimal trajectories have been proposed in the literature, and they can be generally categorized as direct methods (Enright and Conway 1992; Hargraves and Paris 1987) and indirect methods (Jiang et al. 2012; Kechichian 1994; Tang and Jiang 2016). Frameworks that combine direct and indirect methods are usually termed hybrid methods (Gao and Kluever 2004; Kluever and Pierson 1995). In an indirect method, by using the Pontryagin's maximum principle or variation principle, the original optimization problem is transformed into a nonlinear two-point boundary value problem (TPBVP) which is generally solved via shooting methods. The solution from indirect methods is at least locally optimal, since the first-order necessary conditions for optimality are satisfied. However, it is generally difficult for indirect methods to converge to an optimal solution, since the convergence radius of the corresponding TPBVP is small. In addition, such a TPBVP is sensitive to the initial guess for the costate variables, which do not have any intuitive physical meaning. Some effective techniques to overcome the convergence challenge of indirect methods include homotopic transformation (Bertrand and Epenoy 2002; Chi et al. 2017; Haberkorn et al. 2004; Pan et al. 2016), switching detection (Chi et al. 2018; Martinon and Gergaud 2007) and costate variables estimation (Chen et al. 2018; Jiang et al. 2017; Jiang and Tang 2016). For orbit transfer problems with few revolutions, these techniques are proved to be highly efficient. However, when a larger number of orbit revolutions is required for the transfer, indirect methods may struggle in finding convergent solutions. Compared to a TPBVP formulation approach, direct methods transcribe the optimal control problem into a nonlinear programming problem, which generally exhibits a larger convergence domain at the price of increased computational workload. For instance, Betts (2000) uses the direct collocation method paired with sequential quadratic programing to solve a 578-revolution transfer problem, and presents an optimization problem with 416,123 variables and 249,674 constraints. Scheel and Conway (1994) discuss a Runge-Kutta parallel-shooting method for solving a 100-revolution orbit transfer. Solving such large scale optimization problems requires a tremendous computational effort, which put forward higher demand for computational resources. In addition, the solutions that are obtained from direct methods do not generally satisfy the first-order necessary conditions for optimality. Therefore, the converged solutions are not ensured to be locally optimal. In recent years, utilizing convex optimization to solve low-thrust orbit transfer problems has attracted a significant amount of research interest (Tang et al. 2018; Yang et al. 2017a), since it is more computationally tractable compared to nonlinear programming (Liu et al. 2017). It is proved that, convex optimization is highly efficient for solving short-duration trajectory optimization (Yang et al. 2017b). Nevertheless, when it comes to long-duration missions with multiple revolutions, there are no significant advantages from convex optimization. Hybrid methods exhibit both indirect and direct method good properties. The thrust profile is usually assumed a priori, and the optimal control is determined through the optimality conditions that define indirect methods. However, the thrust profile for orbit transfers with multi-revolutions is difficult to be guessed, and therefore, it is difficult to find optimal multi-revolutions solutions with hybrid methods.

Adding to the numerical methods mentioned above, symplectic methods exhibit promising performance in optimal control problems (Peng et al. 2011), owing to the preservation of the symplectic structure of the original problem (Zhong 2006). The symplectic method first convert the nonlinear optimal control problem into a TPBVP using Hamiltonian formulation. Then, based on the dual variational principle, a symplectic form is applied to discretize the TPBVP. After discretization, the optimization problem is described by a set of nonlinear algebraic equations with sparse and symmetric coefficient matrices. Accordingly, solving such type of algebraic equations requires less computational resources. Since the symplectic method is based on the variational principle, it satisfies the first necessary conditions for optimality, which means that the solutions are at least locally optimal. Furthermore, owing to the preservation of the symplectic structure, the symplectic method can yield a reasonable approximation of the continuous solution with fewer discretization points. Peng et al. present a series of symplectic algorithms and utilize them to solve optimal orbit rendezvous problems (Peng et al. 2012), orbit transfer problems between halo orbits (Peng et al. 2014a), optimal nonlinear feedback control for spacecraft rendezvous between libration point orbits (Peng et al. 2014b), bound evaluation for spacecraft swarm reconfiguration on libration point orbits (Peng and Li 2017). Li et al. (2015) introduce the symplectic algorithm with quasi-linearization techniques to solve nonlinear optimal control problems with inequality path constrains, and prove its efficiency for designing spacecraft rendezvous between halo orbits. However, symplectic methods that are presented in existing studies only utilize orbit transfers with one revolution as supporting examples. In addition, the spacecraft mass variation is not taken into consideration in those studies, and should be considered in further research.

The convergence of indirect methods depend on the initial guess for the costates. Compared to indirect methods, the convergence of the symplectic methods mainly depends on the initial guess for the states. Compared to direct methods, symplectic methods require less computational resources, because the final problem formulation incorporates sparse and symmetric coefficient matrices. Consequently, symplectic methods may have large potential for solving optimal control problem with long-duration and multiple revolutions. However, to the authors' best knowledge, no literature has explored the utilization of symplectic algorithms to solve low-thrust orbit transfer problems with many revolutions. That is mainly because, multi-revolution orbit transfers result in oscillation of the state variables through time, which makes difficult for symplectic methods to find convergent solutions. Another reason lies on the fact that, the supporting examples in previous references arbitrarily set the initial guess for the states variables to zero or to a constant value, without providing any reference trajectory. It is proved that, immediately supplying proper reference trajectories can accelerate the convergence of the optimal control problem (Yang et al. 2017c). For those reasons, low-thrust multi-revolution orbit transfers are difficult to optimize via the symplectic method.

In this paper, the application of symplectic algorithms for the optimization of low-thrust orbit transfers with multiple revolutions is investigated in details. An efficient symplectic algorithm is developed to solve the optimal control problem which arises from the original orbit transfer problem. The modified equinoctial elements are applied to describe the motion of the spacecraft. Compared to Cartesian coordinates, the modified equinoctial elements display smaller value oscillations along the final trajectory. A nominal trajectory is given to accelerate the convergence rate of the symplectic method. Three representative low-thrust multi-revolution orbit transfer problems are selected to demonstrate the high accuracy and efficiency of the symplectic algorithm.

This paper is organized as follows. In Section 2, spacecraft dynamics are expressed in modified equinoctial elements, and the model for low-thrust orbit transfer problem is built. The quasi-linearization method is utilized to transcribe the original nonlinear optimal control problem into a sequence of constraint linear-quadratic optimal control sub-problems. In Section 3, a symplectic method is introduced to iteratively solve the sequence of constrained linear-quadratic optimal control sub-problems. To validate the accuracy and efficiency of the symplectic method, three examples of multi-revolution orbit transfer problems are given in Section 4. Concluding remarks are made in Section 5.

## 2 Low-Thrust Orbit Transfer Optimal Control Problem

Consider a transfer problem where the spacecraft is subjected only to gravity of the central body and the thrust of its own electric propulsion system. The objective is to determine the minimum-energy trajectory and thrust vector that transfer the spacecraft from the specified initial states to the specified terminal states. The low-thrust orbit transfer optimal control problem is described next.

### 2.1 Equations of Motion

The state vector consists of the spacecraft position and velocity vectors, which are generally expressed in Cartesian coordinates. However, for low-thrust transfers with a large number of orbit revolutions, Cartesian coordinate values may display strong natural oscillations along the trajectory, which hinder the convergence to an optimal solution. In order to get better convergence performance, this work employs modified equinoctial elements $\boldsymbol{x} = [p, f, g, h, k, L]$ to describe the motion of the spacecraft, where $p$ is the semi-latus rectum of the orbit, and $L$ is the true longitude; the remaining four elements do not have any intuitive physical meaning, however, $f$ together with $g$ can describe the eccentricity of the orbit, and $h$ together with $k$ can describe the inclination of the orbit. Compared to Keplerian orbital elements or Cartesian coordinates, the equinoctial elements are non-singular for most eccentricities and inclinations, except for absolutely retrograde orbit. In addition, equinoctial elements conveniently describe the time variation of the true longitude, which acts as a phase angle. Most important for this work, when the equinoctial elements are chosen to describe the spacecraft motion with multiple revolutions, the natural oscillations of the state variable value can be reduced, and the optimal control problem is easier to solve. The equinoctial elements can be obtained from the Keplerian elements as:

$$\begin{cases} x_1 = p = a(1-e^2) \\ x_2 = f = e\cos(\Omega+\omega) \\ x_3 = g = e\sin(\Omega+\omega) \\ x_4 = h = \tan(i/2)\cos\Omega \\ x_5 = k = \tan(i/2)\sin\Omega \\ x_6 = L = \Omega+\omega+f \end{cases} \quad (1)$$

where $a$ is the semi-major axis, $e$ is the eccentricity of the orbit, $i$ is the inclination of the orbit, $\Omega$ is the longitude of the ascending node, $\omega$ is the argument of perigee, and $\theta$ is the true anomaly. We express the three-dimensional control vector in local vertical/local horizontal (LVLH) coordinates, which are attached to the spacecraft. Then, the spacecraft dynamics can be formulated as follows:

$$\dot{x} = M\left(\frac{T_{max}}{m}u + f_p\right) + D$$
$$\dot{m} = -\frac{T_{max}}{I_{sp}g_0}\|u\| \quad (2)$$

where $M$ is a $6\times 3$ transformation matrix from the LVLH to the equinoctial elements and $D$ is the six-dimensional gravity vector. The expressions of the matrix M and the vector D are as follows:

$$M = \begin{Bmatrix} 0 & \dfrac{2x_1 H}{W} & 0 & 0 & 0 & 0 \\ H\sin x_6 & \dfrac{H}{W}[(W+1)\cos x_6 + x_2] & -\dfrac{HG}{W}x_3 & 0 & 0 & 0 \\ -H\cos x_6 & \dfrac{H}{W}[(W+1)\sin x_6 + x_3] & \dfrac{HG}{W}x_2 & 0 & 0 & 0 \\ 0 & 0 & \dfrac{HS}{2W}\cos x_6 & 0 & 0 & 0 \\ 0 & 0 & \dfrac{HS}{2W}\sin x_6 & 0 & 0 & 0 \\ 0 & 0 & \dfrac{HG}{W} & 0 & 0 & 0 \end{Bmatrix} \quad (3)$$

$$D = [0,0,0,0,0,\dfrac{W^2}{H^3 \mu}] \quad (4)$$

where $\mu$ is the gravitational constant, and the coefficients $H, W, S, G$ are expressed as follows:

$$\begin{aligned} H &= \sqrt{\dfrac{x_1}{\mu}} \\ W &= 1 + x_2 \cos x_6 + x_3 \sin x_6 \\ S &= 1 + x_4^2 + x_5^2 \\ G &= x_4 \sin x_6 - x_5 \cos x_6 \end{aligned} \quad (5)$$

$T_{\max}$ is the maximum thrust magnitude, $m$ is the instantaneous mass of the spacecraft, $g_0$ is the standard gravitational acceleration at sea level, $I_{sp}$ is the specific impulse of the thruster. The control vector is expressed by a three-dimensional vector $u$, with norm between 0 and 1. The symbol $f_p$ represents the perturbation vector. In the central body reference frame, only the central body gravitational force is taken into consideration, and $f_p$ equals $\mathbf{0}$. In the vicinity of Earth, the J2 perturbation is the main perturbation and should be considered. Accordingly, the vector $f_p$ is expressed in LVLH coordinate as follows:

$$f_{pr} = -\frac{3}{2}J_2 \frac{\mu R_e^2 W^4}{x_1^4}\left(1 - \frac{12G^2}{S^2}\right)$$

$$f_{pt} = -\frac{3}{2}J_2 \frac{\mu R_e^2 W^4}{x_1^4}\left\{\frac{4\left[\left(x_4^2 - x_5^2\right)\sin 2x_6 - 2x_4 x_5 \cos 2x_6\right]}{S^2}\right\} \quad (6)$$

$$f_{pn} = -\frac{3}{2}J_2 \frac{\mu R_e^2 W^4}{x_1^4}\left[\frac{4(2-S)G}{S^2}\right]$$

where $R_e$ represents the Earth radius.

In order to facilitate numerical propagation of spacecraft dynamics, the equations of motion are normalized by appropriate characteristic length, time and mass that will be described in Section 4, as they vary with each application. Finally, reference physical constants which will be used in all simulations for this paper, are given in Table 1.

**Table 1** Physical constants

| Quantity | Value |
|---|---|
| $g_{earth}$ | 9.80665 m · s$^{-2}$ |
| $\mu_{earth}$ | 3.9860047 × 10$^{14}$ m$^3$ · s$^{-2}$ |
| $R_e$ | 6,378,140 m |
| $J_2$ | 1082.639 × 10$^{-6}$ |
| $\mu_{sun}$ | 1.327124 × 10$^{20}$ m$^3$ · s$^{-2}$ |

**2.2 Energy-Optimal Control Problem**

An optimal trajectory and control input to transfer the spacecraft from a given orbit state to a target orbit state can be obtained by minimization of energy consumption with appropriate constraint conditions:

$$J = \frac{T_{\max}}{I_{sp}g_0}\int_{t_0}^{t_f}\|\boldsymbol{u}\|^2 \, dt \quad (7)$$

where $t_0$ and $t_f$ denote the initial and final times, respectively, and they are both fixed. In this paper, both rendezvous and orbit transfer problems will be considered. Correspondingly, the boundary conditions for the two scenarios are described as follows:

1. Boundary conditions for rendezvous problems

In rendezvous problems, the initial mass, initial states, and final states are all fixed, while the final mass is free. Thus, the following boundary constraints must be satisfied:

$$x(t_0) = x_0, \quad x(t_f) = x_f \tag{8}$$

$$m(t_0) = m_0, \quad m(t_f) = \text{Free} \tag{9}$$

According to the transversality conditions, the boundary costates are free when the corresponding boundary states are fixed. Thus, the initial costates and the final costates should be free

$$\lambda_x(t_0) = \text{Free}, \quad \lambda_x(t_f) = \text{Free} \tag{10}$$

Since the final mass is free, the final costate of mass should be zero as:

$$\lambda_m(t_f) = 0 \tag{11}$$

2. Boundary conditions for orbit transfer problems

In orbit transfer problems, the initial mass and initial orbit states, are both fixed. In contrast, the final mass is free. In addition, which final states are free or fixed depends on the geometry of the final orbit. If the destination orbit is circular, the following boundary constrains need to be satisfied:

$$\begin{aligned} &x(t_0) = x_0, \\ &p(t_f) = p_0, \quad f(t_f) = 0, \quad g(t_f) = 0 \\ &h(t_f) = \text{Free}, \quad k(t_f) = \text{Free}, \quad L(t_f) = \text{Free} \end{aligned} \tag{12}$$

$$m(t_0) = m_0, \quad m(t_f) = \text{Free} \tag{13}$$

According to the transversality conditions, the initial costates and the final costates should be free or zero as follows:

$$\begin{aligned} &\lambda_x(t_0) = \text{Free} \\ &\lambda_p(t_f) = \lambda_f(t_f) = \lambda_g(t_f) = \text{Free} \\ &\lambda_h(t_f) = \lambda_k(t_f) = \lambda_L(t_f) = 0 \\ &\lambda_m(t_f) = 0 \end{aligned} \tag{14}$$

To maintain the thrust magnitude below its maximum value during the transfer process, the following inequality path constraint is enforced throughout the trajectory:

$$\|u\| \leq 1 \tag{15}$$

The slack variable $\alpha$ is introduced to transform the inequality constraint to an equality form:

$$\|u\| - 1 + \alpha = 0, \quad \alpha \geq 0 \tag{16}$$

Thus, by introducing the costate vector $\lambda = (\lambda_x, \lambda_m)$, which is also known as the functional Lagrange multiplier, and the parameter variable $\beta$, the Hamilton function can be defined as follows:

$$H = \lambda_x \cdot [M(\frac{T_{\max}}{m}u + f_p) + D] - \lambda_m \cdot \frac{T_{\max}}{I_{sp}g_0}\|u\| + \frac{T_{\max}}{I_{sp}g_0}\|u\|^2 + \beta^T(\|u\| - 1 + \alpha) \tag{17}$$

Moreover, an augmented cost function can be obtained:

$$J_A = \int_0^{t_f} [H - \lambda^T \dot{x}] dt \tag{18}$$

After computing the variations of the augmented cost function, the optimal solutions should satisfy the following Hamiltonian canonical equations:

$$\dot{x} = \frac{\partial H}{\partial \lambda} \tag{19}$$

$$\dot{\lambda} = -\frac{\partial H}{\partial x} \tag{20}$$

The first order necessary conditions of the optimal control problem can be obtained by the following equations:

$$\frac{\partial H}{\partial u} = 0 \tag{21}$$

$$\frac{\partial H}{\partial \beta} = 0 \tag{22}$$

According to the Pontryagin's minimum principle, the following complementary conditions can be derived:

$$\alpha \geq 0, \ \beta \geq 0, \ \alpha^T \beta = 0 \tag{23}$$

Thus, the optimal solution can be obtained by solving Eqs. (19)-(23) with boundary conditions Eqs. (8)-(11) or Eqs. (12)-(14).

**2.3 Quasilinearizaiton Method**

In order to construct symplectic-preserving condition, quasilinearization techniques are applied in this paper. The state and constraint equations are linearized, while the cost function is expanded up to second order around a reference solution. The solution to the quasilinear problem is, then, utilized as new reference, and this process is iterated until convergence. Each time the reference is updated, the algorithm advances by one iteration. Thus, the original nonlinear optimal control problem is transformed into a sequence of constrained linear quadratic optimal control sub-problems that can be solved individually via a symplectic method.

Denoting the state vector $x = (p, f, g, h, k, L, m)$ and the control vector $u = (u_x, u_y, u_z)$, the constrained linear quadratic optimal control sub-problem at the $(k+1)$ iteration can be described by the following state equations:

$$\dot{x}^{(k+1)} = A^{(k)}x^{(k+1)} + B^{(k)}u^{(k+1)} + w^{(k)} \tag{24}$$

where

$$A^{(k)} = \frac{\partial f(x,u,t)}{\partial x}\bigg|_{x^{(k)},u^{(k)}}, B^{(k)} = \frac{\partial f(x,u,t)}{\partial u}\bigg|_{x^{(k)},u^{(k)}} \tag{25}$$

$$w^{(k)} = f(x^{(k)},u^{(k)},t) - A^{(k)}x^{(k)} - B^{(k)}u^{(k)} \tag{26}$$

Subject to the path constraints:

$$h^{(k+1)}(x,u,t) = C^{(k)}x^{(k+1)} + D^{(k)}u^{(k+1)} + v^{(k)} \leq 0 \tag{27}$$

where

$$C^{(k)} = \frac{\partial h(x,u,t)}{\partial x}\bigg|_{x^{(k)},u^{(k)}}, D^{(k)} = \frac{\partial h(x,u,t)}{\partial u}\bigg|_{x^{(k)},u^{(k)}} \tag{28}$$

$$v^{(k)} = h(x^{(k)},u^{(k)},t) - C^{(k)}x^{(k)} - D^{(k)}u^{(k)} \tag{29}$$

The cost function is also transformed into:

$$J^{(k+1)} = \int_{t_0}^{t_f} g^{(k+1)} \mathrm{d}t \tag{30}$$

where

$$g^{(k+1)} = \bar{g}^{(k)} + (u^{(k+1)} - u^{(k)})E^{(k)} + \frac{1}{2}(u^{(k+1)} - u^{(k)})^T F^{(k)}(u^{(k+1)} - u^{(k)}) \tag{31}$$

$$\bar{g}^{(k)} = \frac{1}{2}\|u\|^2 \bigg|_{u^{(k)}} \tag{32}$$

$$E^{(k)} = \frac{\partial \bar{g}}{\partial u}\bigg|_{u^{(k)}}, F^{(k)} = \frac{\partial^2 \bar{g}}{\partial u^2}\bigg|_{u^{(k)}} \tag{33}$$

Superscript in the above equations are an iteration index: the symbol $(k+1)$ denotes variable values in the current $(k+1)$ iteration, $(k)$ refers to values at the previous $k$ iteration, which serve as the initial reference for the current update.

Therefore, the original nonlinear optimal control problem is transformed into a sequence of constrained linear quadratic control sub-problems. The iteration process ends when the variation of the orbit states is smaller than a given tolerance. The convergence criteria is defined

as:

$$\frac{\|x_{k+1} - x_k\|}{\|x_k\|} \leq \varepsilon \quad (34)$$

where $\varepsilon$ is a small quantity which denotes the selected tolerance. Next, a symplectic method is proposed to obtain the solution of the linear quadratic control sub-problem at each iteration.

## 3 Symplectic Approach for Constrained Linear Quadratic Optimal Control

A symplectic method based on dual variational principle is proposed to obtain the solution of the linear quadratic optimal control sub-problems. First, the linear quadratic optimal control problem is converted into a series of nonlinear algebraic equations by using a symplectic method in discrete form. Then, based on complementary conditions in Eq. (23), the explicit linear complementary problem can be derived. For brevity, the iteration index will be omitted in the following derivations.

### 3.1 Construction of the Symplectic Approach

First, the construction of the symplectic approach is introduced. The derivation follows that in (Peng et al. 2011). The Hamilton function for each constrained linear quadratic optimal control sub-problem can be obtained as follows:

$$H = \bar{g} + (u - u_d)E + \frac{1}{2}(u - u_d)^T F(u - u_d) + \lambda(Ax + Bu + w) + \beta(Cx + Du + v + \alpha) \quad (35)$$

Where the subscript $d$ represents the initial reference for the current iteration. Substituting Eq. (35) into Eq. (21), the expression for the control variable at the $(k+1)$ iteration can be obtained as:

$$u = u_d - F[E + B^T \lambda + D^T \beta] \quad (36)$$

Substituting Eq. (36) back into Eq. (35), yields

$$H = \bar{g} - \frac{1}{2}(E^T + \lambda^T B + \beta^T D)F^{-1}(E + B^T \lambda + D^T \beta) \\ + \lambda(Ax + w) + \beta(Cx + v + \alpha) + (\lambda B + \beta D)u_d \quad (37)$$

In addition, by substituting Eq. (36) back into Eq. (29), the Eq. (16) can be rewritten as

$$Cx + Du_d - DF^{-1}\left[E + B^T \lambda + D^T \beta\right] + v + \alpha = 0 \quad (38)$$

Therefore, the Hamilton function is independent from the control variable. We define an action $S$ in a generic time interval $(a,b)$ as

$$S = \int_a^b [\lambda^T \dot{x} - H] dt \quad (39)$$

Based on the action $S$, the fourth kind of generating function is produced:

$$V = \lambda_a^T x_a - \lambda_b^T x_b + S \quad (40)$$

Variation of the fourth generating function yields:

$$\delta V = x_a^T d\lambda_a - x_b^T d\lambda_b + \int_a^b [(\dot{x} - \frac{\partial H}{\partial \lambda})\delta\lambda - (\dot{\lambda} + \frac{\partial H}{\partial x})\delta x] dt \quad (41)$$

According to the variation principle, the optimal solution should satisfy the Hamilton canonical equations.

$$\delta V = x_a^T d\lambda_a - x_b^T d\lambda_b \quad (42)$$

In this formulation, the costate variables at the extremes of the time interval $(a,b)$ are the free variables, also called independent variables. It can be demonstrated that, numerical method which satisfied Eq. (42) can be symplectic-preserving referring to .

Next, the trajectory is divided into $N$ arcs with equal time intervals $\eta = (t_f - t_0)/N$. The costate variables at both ends of each arc form the set of independent variables. Within each trajectory arc, the state vector $x(t)$ and the costate vector $\lambda(t)$ are approximated by using

Lagrange interpolation polynomials with order $m-1$ and $n-1$, respectively; note that, state and costate variables at the internal interpolation points are not considered independent variables. The parameter variables $\alpha$ and $\beta$ are assumed to be constant. The resulting system of expressions is

$$x(t) = (M \otimes I)\bar{x}_j \tag{43}$$

$$\lambda(t) = N_1 \lambda_{j-1} + (\underline{N} \otimes I)\bar{\lambda}_j + N_n \lambda_j \tag{44}$$

$$\alpha(t) = \alpha_j \tag{45}$$

$$\beta(t) = \beta_j \tag{46}$$

where $\bar{x}_j$ comprises all the state variables at both the extreme and interpolation points within the $j$th arc, defined as $\bar{x}_j = [\bar{x}_j^1, \bar{x}_j^2, ..., \bar{x}_j^m]^T$, $\lambda_{j-1}$ and $\lambda_j$ denote the costate variables at the left and right end of the $j$th arc, $\bar{\lambda}_j$ is composed of the remaining dependent costate variables at the interpolation points within the $j$th arc, defined as $\bar{\lambda}_j = [\bar{\lambda}_j^2, \bar{\lambda}_j^3, ..., \bar{\lambda}_j^{n-1}]^T$, $I$ denotes a $n \times n$ identity matrix, and the symbol $\otimes$ denotes the Kronecker product of two matrices. A scheme of the trajectory discretization is depicted in Figure 1. Other matrices in Eq. (43) and Eq. (44) are defined as

$$M = [M_1, M_2, ..., M_m] \tag{47}$$

$$\underline{N} = [N_2, N_3, ..., N_{n-1}] \tag{48}$$

$$M_i = \prod_{j=1, j \neq i}^{m} \frac{t - (j-1)\eta/(m-1)}{(i-j)\eta/(m-1)} \tag{49}$$

$$N_i = \prod_{j=1, j \neq i}^{n} \frac{t - (j-1)\eta/(n-1)}{(i-j)\eta/(n-1)} \tag{50}$$

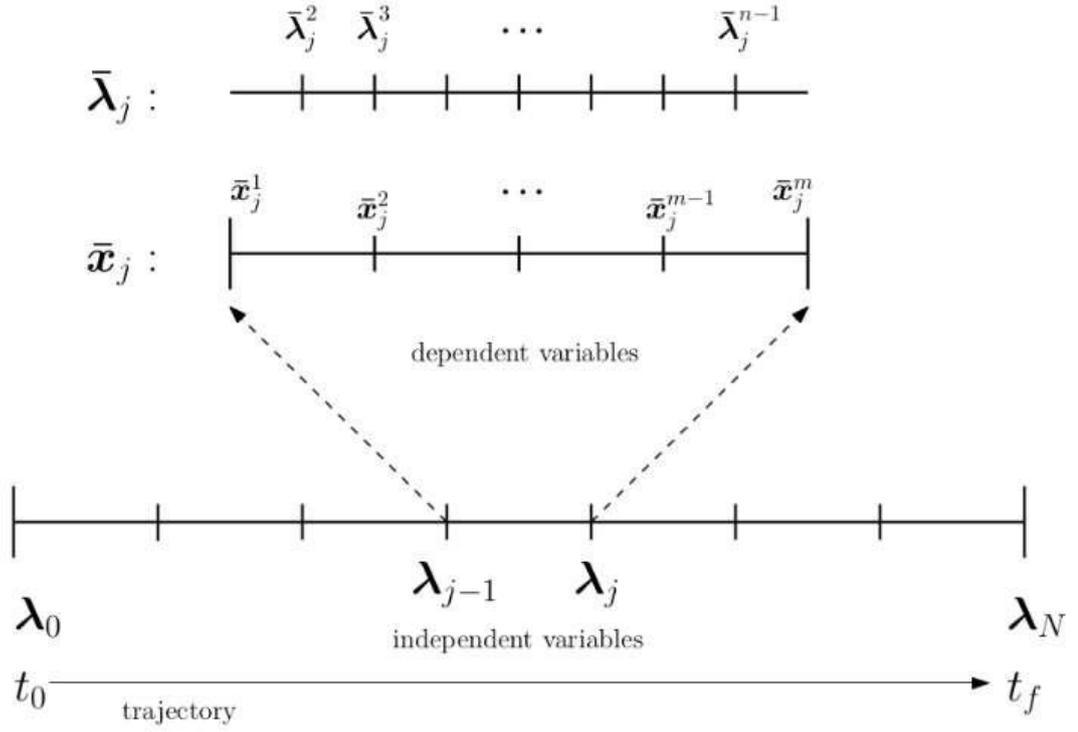

**Fig. 1** Trajectory discretization scheme

Substituting interpolated state and costate variables Eq. (43) and Eq. (44) into Eq. (41) gives

$$V_j = \lambda_{j-1}^T \bar{x}_j^1 - \lambda_j^T \bar{x}_j^r + \int_{t_{j-1}}^{t_j} (\lambda^T \dot{x} - H) \mathrm{d}t \tag{51}$$

Hence, based on Eq. (42), the following equations must be satisfied at each arc:

$$F_1^j = x_{j-1} \tag{52}$$

$$F_2^j = \mathbf{0} \tag{53}$$

$$F_3^j = \mathbf{0} \tag{54}$$

$$F_4^j + x_j = \mathbf{0} \tag{55}$$

where

$$F_1^j = \frac{\partial V_j}{\partial \lambda_{j-1}} = K_{11}^j \lambda_{j-1} + \left( E_u^\mathrm{T} + K_{12}^j \right) \bar{x}_j + K_{13}^j \bar{\lambda}_j + K_{14}^j \lambda_j + f_1^j \tag{56}$$

$$F_2^j = \frac{\partial V_j}{\partial \bar{x}_j} = \left(K_{21}^j + E_u\right)\lambda_{j-1} - K_{22}^j \bar{x}_j + K_{23}^j \bar{\lambda}_j + \left(K_{24}^j - E_d\right)\lambda_j + f_2^j \tag{57}$$

$$F_3^j = \frac{\partial V_j}{\partial \bar{\lambda}_j} = K_{31}^j \lambda_{j-1} + K_{32}^j \bar{x}_j + K_{33}^j \bar{\lambda}_j + K_{34}^j \lambda_j + f_3^j \tag{58}$$

$$F_4^j = \frac{\partial V_j}{\partial \lambda_j} = K_{41}^j \lambda_{j-1} + \left(K_{42}^j - E_d^{\mathrm{T}}\right)\bar{x}_j + K_{43}^j \bar{\lambda}_j + K_{44}^j \lambda_j + f_4^j \tag{59}$$

The detailed expressions for $K_{u,v}^j (u,v=1,2,3,4)$ and $f_i^j (i=1,2,3,4)$ in Eq. (56)-(59) can be found in (Peng et al. 2014b).

### 3.2 Formulation of the Complementary Problem

The $\bar{x}_j$ and $\bar{\lambda}_j$ vectors can be expressed using the independent variables $\lambda_{j-1}$ and $\lambda_j$ by solving Eq. (53) and Eq. (54). After that, substituting the expression for vectors $\bar{x}_j$ and $\bar{\lambda}_j$ into Eq. (52) and Eq. (55), yields

$$F_1^j = S_{11}^j \lambda_{j-1} + S_{12}^j \lambda_j + \zeta_1^j \tag{60}$$

$$F_4^j = S_{21}^j \lambda_{j-1} + S_{22}^j \lambda_j + \zeta_2^j \tag{61}$$

The detailed expression of $S_{u,v}^j (u,v=1,2)$ and $\zeta_i^j (i=1,2)$ can be seen in (Peng et al. 2014b) (see Eq. (50-57)). The $\zeta_1^j$ and $\zeta_2^j$ vectors can be rewritten as

$$\zeta_1^j = \zeta_{11}^j + \zeta_{12}^j \beta_j \tag{62}$$

$$\zeta_2^j = \zeta_{21}^j + \zeta_{22}^j \beta_j \tag{63}$$

The detailed expressions for the above $\zeta_{pq}^j (p,q=1,2)$ vector can also be found in (Peng et al. 2014b) (see Appendix).

Applying Eq. (42) to the each arc throughout the entire trajectory, yields the following nonlinear equation:

$$F_1^1 = x_0 \tag{64}$$

$$F_1^{j+1} + F_4^j = \mathbf{0} \left( j = 1, \cdots, N-1 \right) \tag{65}$$

$$F_4^N + x_f = \mathbf{0} \tag{66}$$

By utilizing Eq. (60)-(66), the costate variable $\hat{\lambda} = \left\{ \lambda_0^T, \lambda_1^T, ..., \lambda_j^T, ..., \lambda_N^T \right\}$ can be expressed in the parameter variable $\hat{\beta} = \left\{ \beta_1^T, \beta_2^T, ..., \beta_j^T, ..., \beta_N^T \right\}$

$$\hat{\lambda} = A^{-1} \Psi \hat{\beta} + A^{-1} \Phi \tag{67}$$

Refer to (Li et al. 2015) for a detailed expressions for the above matrices $A, \Psi, \Phi$. Then, the state vector can be expressed in the parameter $\hat{\beta}$ by utilizing Eq. (55) and Eq. (61)

$$x_j = -F_4^j = -(S_{21}^j \lambda_{j-1} + S_{22}^j \lambda_j + \zeta_2^j) \tag{68}$$

That allows to express the state and costate vectors at the discretization points along the trajectory through the parameter variable $\hat{\beta}$. By discretization points we refer to the collection that comprises both interpolation and independent points. Moreover, the complementary conditions in Eq. (23) also need to be satisfied. The complementary conditions along with the inequality constraints Eq. (38) are enforced at the discretization points; thus, a standard linear complementary problem can be obtained:

$$\hat{\alpha} - M_{new} \hat{\beta} = q_{new} \tag{69}$$

$$\hat{\alpha} \geq \mathbf{0}, \quad \hat{\beta} \geq \mathbf{0} \tag{70}$$

$$\hat{\beta}^T \hat{\alpha} = \mathbf{0} \tag{71}$$

The symbol $M_{new}, q_{new}$ and derivation process follows that in (Li et al. 2015). In general, the parameter $\hat{\beta}$ at the discretization points can be obtained by solving the standard linear complementary problem. Next, the costate vector can be obtained by substituting $\hat{\beta}$ into Eq. (67); similarly the state vector is produced by Eq. (68). Finally, the control input is derived by Eq. (36). Following this procedure, constrained linear quadratic optimal control with given terminal states may be solved. In addition, the matrices $A, \Psi, \Phi, M_{new}$ all have sparse structure with small band width, which makes the numerical implementation of the proposed method highly efficient. It should be noted that, Eq. (67) and Eq. (69) can be modified to reflect the desired boundary conditions.

### 3.3 Treatment of the boundary conditions

As it is described in Section 2.2, boundary conditions for rendezvous problems and orbit transfer problems are considered in this paper. In the case of boundary conditions for rendezvous problems, the costate for the final mass equals zero. Thus, the last row of vector $\lambda_N$ is removed from the list of unknown variables. Correspondingly, the last row of the $\hat{\beta}$, Eq. (67), Eq. (69) should also be deleted. In the case of boundary conditions for orbit transfer problems, the costate of the final mass and the costate of the last three components of the state vector are zero. Thus, the last four rows of $\lambda_N$ are removed from the list of unknown variables. Similarly, the last four rows of $\hat{\beta}$, Eq. (67), Eq. (69) should be deleted.

### 4 Numerical Examples and Discussions

This section presents three examples of energy-optimal transfers with multiple revolutions to illustrate the accuracy and efficiency of the techniques and methods presented in last two sections. To capture the oscillation of the state variables well, the number of the interpolation points for the state variables in a sub-interval is set to be 4, and that for the costate variables is set to be 5. All computations are performed in MATLAB (R2016b) on a desktop computer with a CPU of 4.00 GHz. The heliocentric position and velocity vectors of the planets, when needed, are computed online using the Jet Propulsion Laboratory Horizons system.[3] The value of convergence tolerance in Eq. (34) is set to be 1.0e-6.

**4.1 Generation of Nominal Trajectories**

Since the state equations of spacecraft are linearized around a sequence of reference trajectories, the iteration process for achieving an acceptable error is impacted by the quality of the initial guess, especially for orbit transfer problems with multiple revolutions. If the initial nominal trajectory is too far from the true optimal trajectory, the symplectic method presented in this paper may not converge to the optimal solution. In contrast to other studies, the initial reference trajectory is generated by linear interpolation of the state variables, which are the non-singular equinoctial orbital elements and the mass of the spacecraft between the initial and final trajectory points. Empirically, that results in an effective strategy for multi-revolution transfers. Initially, the control variable value at every discretization point is identically set to 0.005N. Although the initial nominal trajectory may be neither optimal nor feasible, an optimal, feasible trajectory can be generally obtained after a small number of iterations with the symplectic method.

**4.2 Rendezvous from Earth to Venus**

A low-thrust rendezvous problem from Earth to Venus is considered in this section. Namely, the spacecraft starts at the instantaneous Earth heliocentric position and velocity and arrives at

---

[3] Data available online at http://ssd.jpl.nasa.gov/?horizons

Venus with its same instantaneous heliocentric position and velocity. This example exactly replicates that in (Jiang et al. 2012), and is presented to illustrate the accuracy of results that are obtained by our optimization strategy. In (Jiang et al. 2012) the optimal trajectory is obtained via an indirect method and will serve as comparison. The method in (Jiang et al. 2012) include an homotopic transformation of the solution. Since we search for the energy-optimal trajectory we only consider the solution for an homotopic parameter equals to one. All the input parameters are listed in Table 2. For computational convenience, length, time, and mass are nondimensionalized by the astronomical unit (AU, 149597870.66 km), solar year (yr, $365.25 \times 86,400$ s), and spacecraft initial mass, respectively.

Table 2 Parameters for a representative Earth-to-Venus transfer

| Parameter | Value | Units |
|---|---|---|
| Initial date | 7 Oct. 2005 0:0:0:0 | Coordinate Time |
| Flight time | 1000 | Day |
| Initial position | [$9.708322 \times 10^{-1}, 2.375844 \times 10^{-1}, -1.671055 \times 10^{-6}$] | AU |
| Initial velocity | [$-1.598191, 6.081958, 9.443368 \times 10^{-5}$] | AU/yr |
| Final Position | [$-3.277178 \times 10^{-1}, 6.389172 \times 10^{-1}, 2.765929 \times 10^{-2}$] | AU |
| Final Velocity | [$-6.598211, -3.412933, 3.340902 \times 10^{-1}$] | AU/yr |
| $I_{sp}$ | 3800 | s |
| $T_{max}$ | 0.33 | N |
| $m_0$ | 1500.0 | kg |

Converged results obtained by the symplectic method with different number of trajectory arcs are listed in Table 3. It can be seen in Table 3 that, the number of iterations to converge to the optimal solution is not influenced by the number of arcs. In contrast, a dozen of grams may add to the final mass if the number of arcs is increased. We also note from Table 3 that, the third decimal place of the final mass value is converged when the number of arcs equals 20, while the fourth decimal place of the final mass value is converged when the number of arcs equal 30. The indirect method predicts a final mass of 1274.956883kg, with a variation of 0.0304kg when the number of arcs equals 10, and a variation of 0.00043kg when the number of arcs equals 35. Thus, the relationship between the number of arcs and the accuracy of the optimal solution can be inferred. That is, if one revolution contains 3 or 4 arcs, the symplectic method can produce the optimal solution with relatively high accuracy. When the number of arcs equal to 9 or 10 in a revolution, the symplectic method can achieve the same precision of the indirect method. This fact is also demonstrated in other numerical examples. Furthermore, it is concluded that the accuracy of the proposed method can be improved by increasing the number of discretization points. Since solutions via the indirect method are guaranteed to be at least locally optimal, the optimality of the trajectories produced by the symplectic method is also demonstrated in this example.

**Table 3** Converged solutions from the symplectic method with different number of arcs

| Number of Arcs | Number of Iterations | Computational Time(s) | Final Mass(kg) |
| --- | --- | --- | --- |
| 5 | 9 | 0.136186 | 1274.747782 |
| 10 | 9 | 0.326455 | 1274.959963 |
| 15 | 9 | 0.363343 | 1274.959691 |
| 20 | 9 | 1.220348 | 1274.957674 |
| 25 | 9 | 1.749006 | 1274.957120 |
| 30 | 9 | 3.145607 | 1274.956974 |
| 35 | 9 | 4.625240 | 1274.956923 |

A comparison of the optimal low-thrust multi-revolution trajectories solved by the symplectic method and indirect method is displayed in Figure 2. The symplectic method uses 35 arcs. Parameters of the indirect method are set to follow (Jiang et al. 2012). The symbol "*" represents the trajectory obtained from the symplectic method, while the dashed orange line renders the trajectory obtained from the indirect method. Figure 2 portrays the path of the spacecraft from the initial Earth heliocentric position and velocity to the Venus rendezvous by matching Venus heliocentric position and velocity after 3 orbital revolutions. Both the symplectic method and the indirect method produce nearly identical optimal low-thrust trajectories.

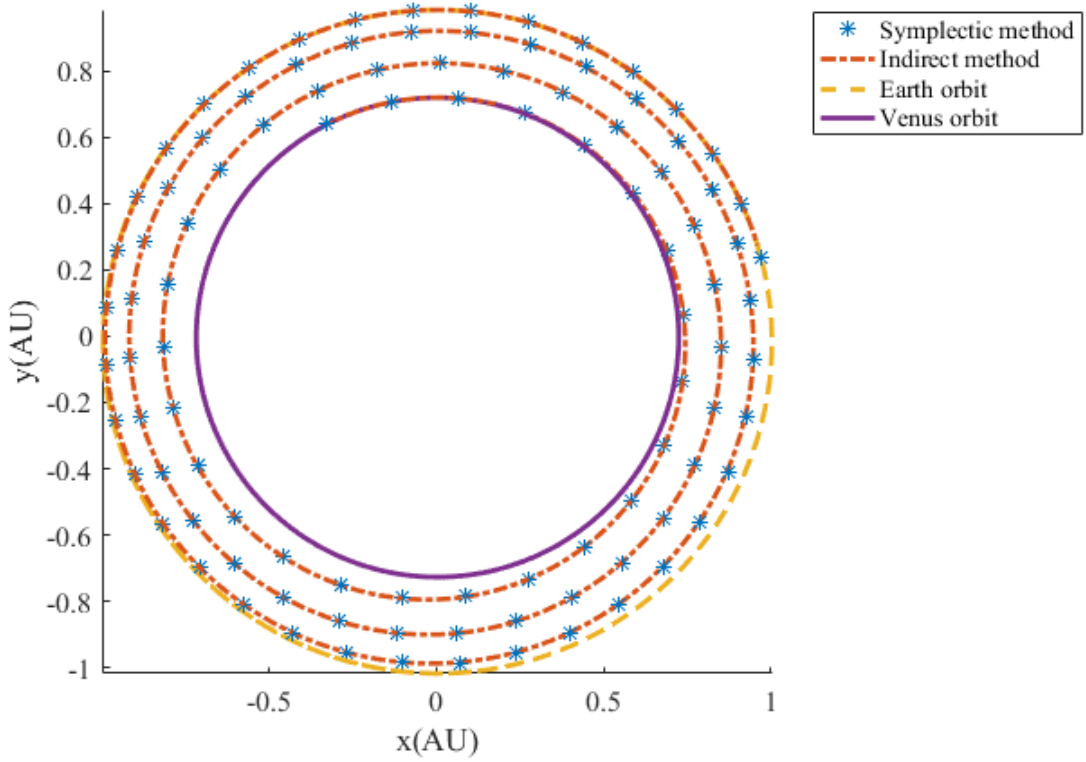

**Fig. 2** Low-thrust trajectory from Earth to Venus

The time evolution for costate variables of the symplectic and indirect method is depicted in Figure 3, denoted by stars and lines respectively. From Figure 3, it is clear that the costate variables obtained from the two methods are in a close agreement. From Figure 3 it is easy to verify that the terminal mass costate $\lambda_m$ satisfies the transversallity condition in Eq. (14), i.e. $\lambda_m(t_f)=0$, and demonstrates that boundary conditions for the spacecraft mass costate are also satisfied.

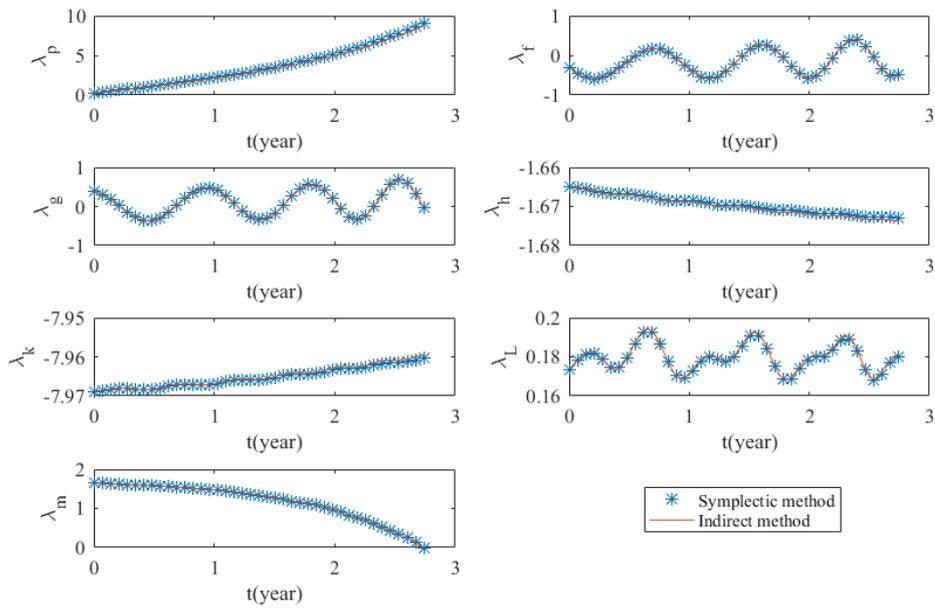

**Fig. 3** Costate variables time histories for the optimal rendezvous trajectory from Earth to Venus

The optimal thrust profile is displayed in Figure 4. Both, the symplectic and indirect method converge on nearly identical optimal thrust profiles. In addition, the thrust magnitude is below one during the entire transfer, and the path constraint in Eq. (15) is satisfied.

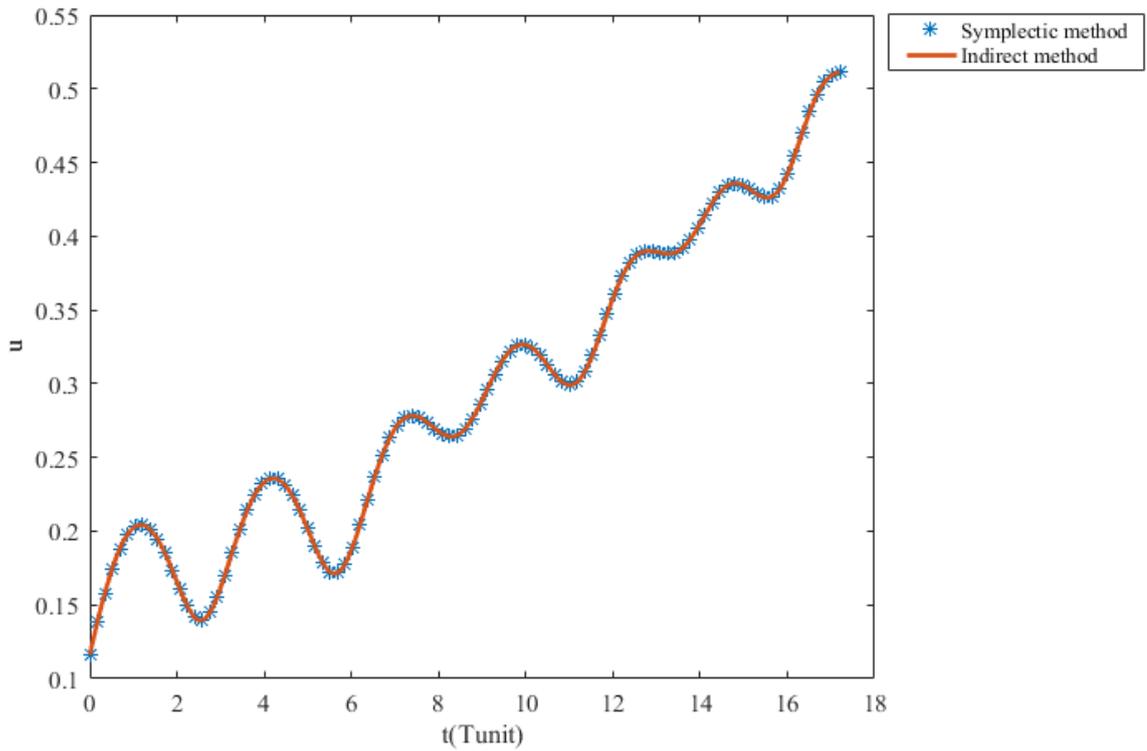

**Fig. 4** Optimal thrust profile of the rendezvous from Earth to Venus

Thus, for this problem, since the solution from the two methods are in close agreement, we can conclude that the symplectic method converges on the locally optimal solution with high accuracy.

**4.3 Orbital Transfer between Two Circular Orbits**

In this section, a low-thrust orbit transfer problem between two circular orbits around the Sun is considered: the spacecraft starts from the instantaneous Earth heliocentric position and velocity and arrives at a final, given circular orbit. This example replicates that in (Lantoine and Russell 2012), and is presented to illustrate the efficiency of the symplectic method. The specific impulse $I_{sp}$ is assumed to be constant and equals to 2000 s and the initial mass of the spacecraft is 1000 kg. The initial epoch is 00:00:00, April 10th, 2007, and the corresponding Earth position and velocity vectors at this epoch are retrieved from JPL ephemerides DE405:

$$\boldsymbol{r}_0 = [-140699693, -51614428, 980]\,\text{km} \tag{72}$$

$$\boldsymbol{v}_0 = [9.774596, -28.07828, 4.337725 \times 10^{-4}]\,\text{km/s} \tag{73}$$

Which can be transformed into equinoctial orbit elements:

$$\begin{aligned}
p(t_0) &= 0.999725801184726 \text{ AU} \\
f(t_0) &= -0.003755794501262 \\
g(t_0) &= 0.016268822901105 \\
h(t_0) &= -0.000007924683518 \\
k(t_0) &= 0.000000575495165 \\
L(t_0) &= 3.493191186522740 \text{ Rad}
\end{aligned} \tag{74}$$

The final spacecraft terminates in a circular orbit with radius $a_{\text{target}} = 1.95\text{AU}$. Since the final orbit is a circular orbit, the eccentricity is zero. The remaining four Keplerian elements are free. The corresponding equinoctial orbit elements are:

$$\begin{aligned}
p(t_f) &= 1.95 \text{ AU} \\
f(t_f) &= 0 \\
g(t_f) &= 0 \\
h(t_f) &= \text{Free} \\
k(t_f) &= \text{Free} \\
L(t_f) &= \text{Free}
\end{aligned} \tag{75}$$

To facilitate numerical computations, length, time, and mass are nondimensionalized as in the last section. Since in both the current and previous example, the central body is the Sun, both problems can be nondimensionalized by the same characteristic quantities.

To better understand the influence of the number of revolutions on the optimization process, the optimal transfer is solved for a set of four different times of flight (which correspond to a different number of revolutions). For each given time of flight, the maximum thrust magnitude is adjusted to ensure that there exists a feasible low-thrust transfer. Resulting parameters for the four time of flight cases are listed in Table 4.

**Table 4** Parameters for the numerical examples

| Case | $N_{rev}$ | $T_{max}$(N) | TOF(days) |
|------|-----------|--------------|-----------|
| 1    | 2         | 0.2          | 1165.65   |
| 2    | 5         | 0.14         | 2325.30   |
| 3    | 9         | 0.05         | 4650.60   |
| 4    | 17        | 0.015        | 8719.88   |

The results produced by the optimal control software GPOPS are chosen for comparison to illustrate the efficiency of the symplectic method. GPOPS is an open source MATLAB software developed by Rao et al. for solving complex optimal control problems using the nonlinear programming solver SNOPT, which is developed by Gill. To make a legit comparison of the algorithm efficiency, the initial guess of state variables and control variables are set the same for both the symplectic method and the GPOPS. As for the other parameters of GPOPS, they are listed in the Table 5.

**Table 5** Parameters for the optimal control software GPOPS

| Parameters | Value |
|------------|-------|
| setup.mesh.tolerance | 1e$^{-6}$ |
| setup.mesh.iteration | 30 |
| setup.derivatives | finite-difference |
| setup.checkDerivatives | 0 |
| setup.autoscale | off |

The results for different time of flight cases are listed in Table 6. From Table 6, it can be found that, the final mass obtained by symplectic methods is nearly the same as that of GPOPS, which means the solution produced by symplectic methods can have the same accuracy as that of GPOPS. Moreover, it should be noted that, the symplectic method converges to the optimal solution with fewer discretization points when compared to GPOPS. Thus, it can be concluded that the symplectic method can preserve the continuous nature of the original dynamics with fewer discretization points when compared to GPOPS. Besides, the symplectic method is significantly faster than GPOPS in terms of total computational time. Thus, for this example, the efficiency and optimality of the symplectic method can be demonstrated.

**Table 6** Comparison between the symplectic method and SNOPT solver for multi-revolution orbital transfers

| Case | Method | $m_f$ (kg) | Number of discretization points | CPU time (s) |
|---|---|---|---|---|
| 1 | Symplectic method | 647.5883 | 40 | 0.2269 |
|   | GPOPS/SNOPT | 647.5883 | 321 | 2.2056 |
| 2 | Symplectic method | 649.1790 | 60 | 0.3747 |
|   | GPOPS/SNOPT | 649.1790 | 466 | 3.7903 |
| 3 | Symplectic method | 649.6894 | 80 | 0.4891 |
|   | GPOPS/SNOPT | 649.6878 | 897 | 8.0261 |
| 4 | Symplectic method | 649.6167 | 200 | 3.9055 |
|   | GPOPS/SNOPT | 649.6168 | 1661 | 38.1206 |

For reference, the optimal solution obtained by symplectic method for Case 3 is depicted in Figures. 5, 6, 7, 8. Figure 5 shows the optimal spacecraft trajectory obtained from the symplectic method and SNOPT solver. The complete orbit transfer contains nearly 9 revolutions. The thrust profiles are portrayed in Figure 6. From Figure 5 and Figure 6, it can be noticed that, the optimal transfers obtained from the two methods are in close agreement. The evolution of the final mass and final true anomaly with the number of iterations during the optimization process for the symplectic method is depicted in Figure 7 and Figure 8. The final mass approximately converges after 3 iterations. During the remaining iterations, the solution slowly updates its final true longitude. As a consequence, the efficiency of the symplectic method can be further improved by giving better initial guesses for the true longitude.

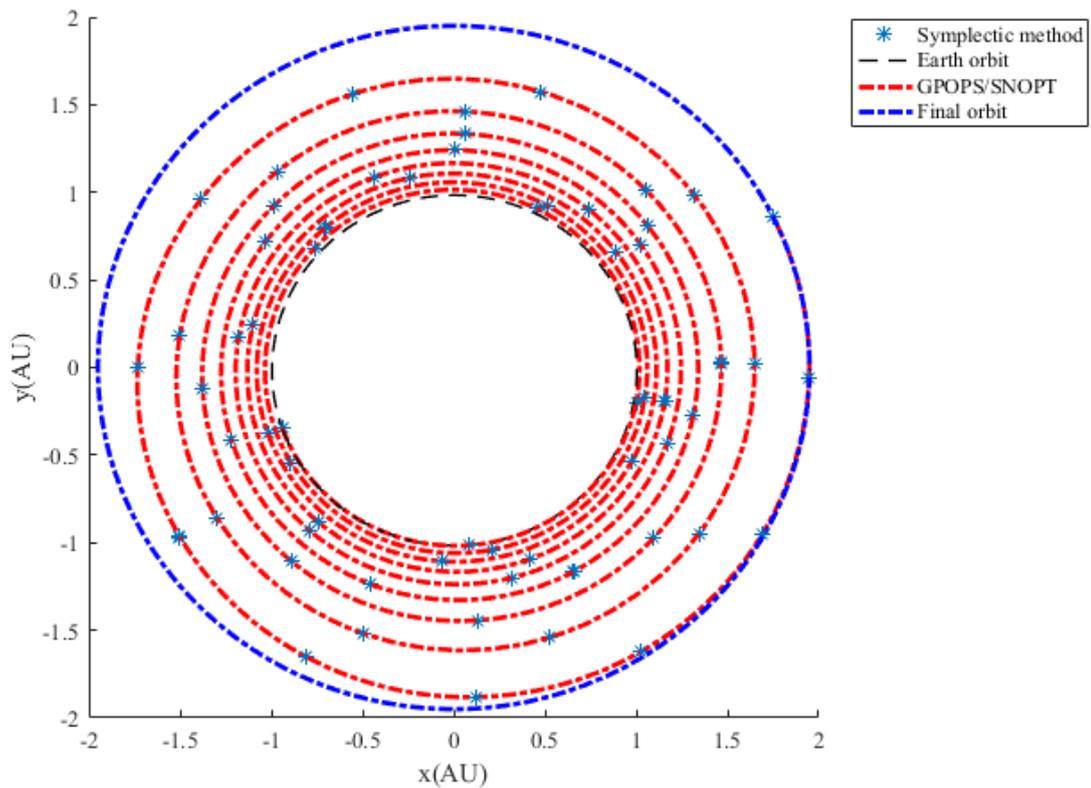

**Fig. 5** Low-thrust trajectory from Earth to a circular orbit

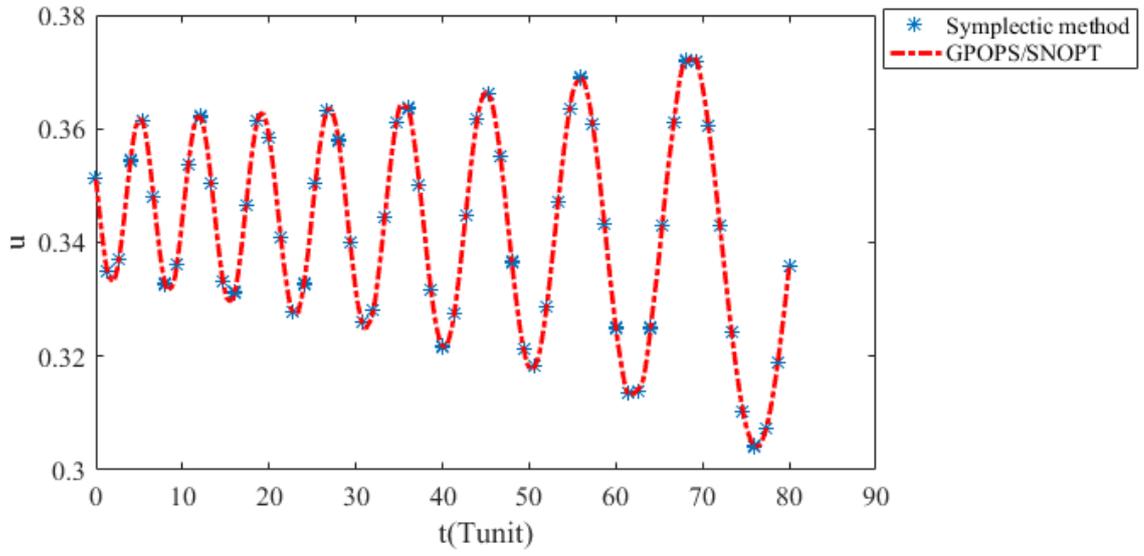

**Fig. 6** Optimal thrust profile of the orbit transfer from Earth to a circular orbit

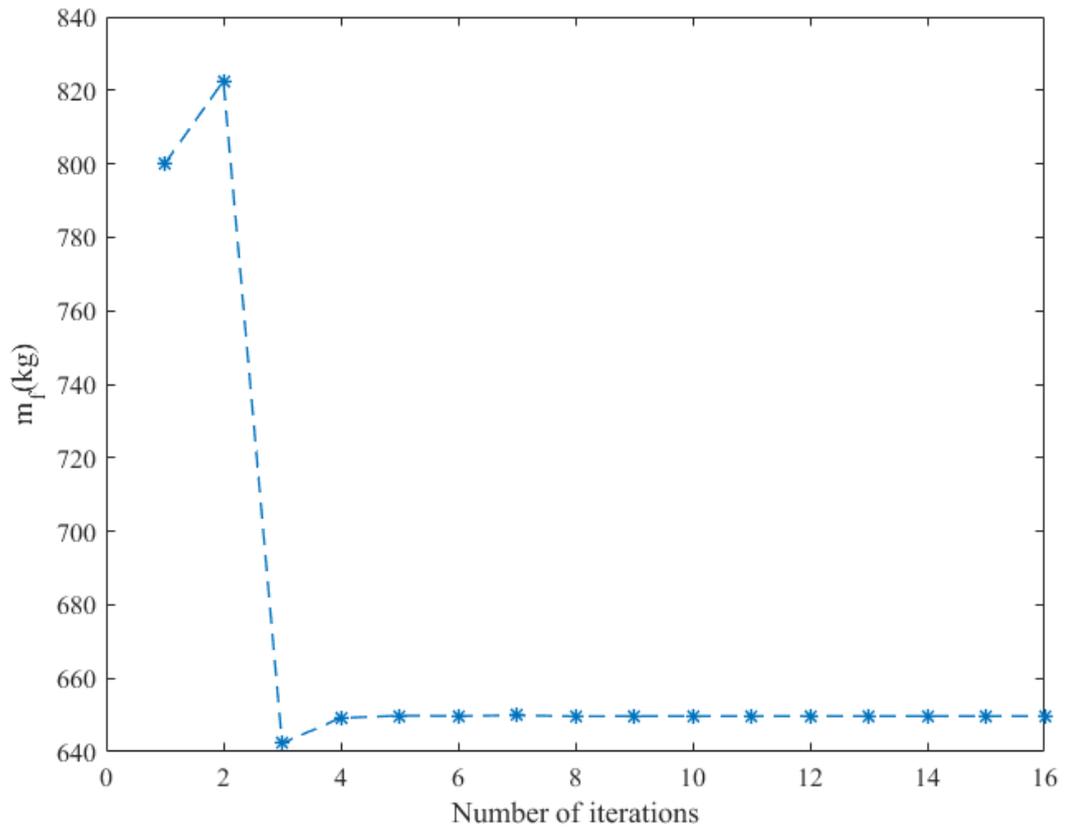

**Fig. 7** Evolution of the final mass during the optimization process for the symplectic method

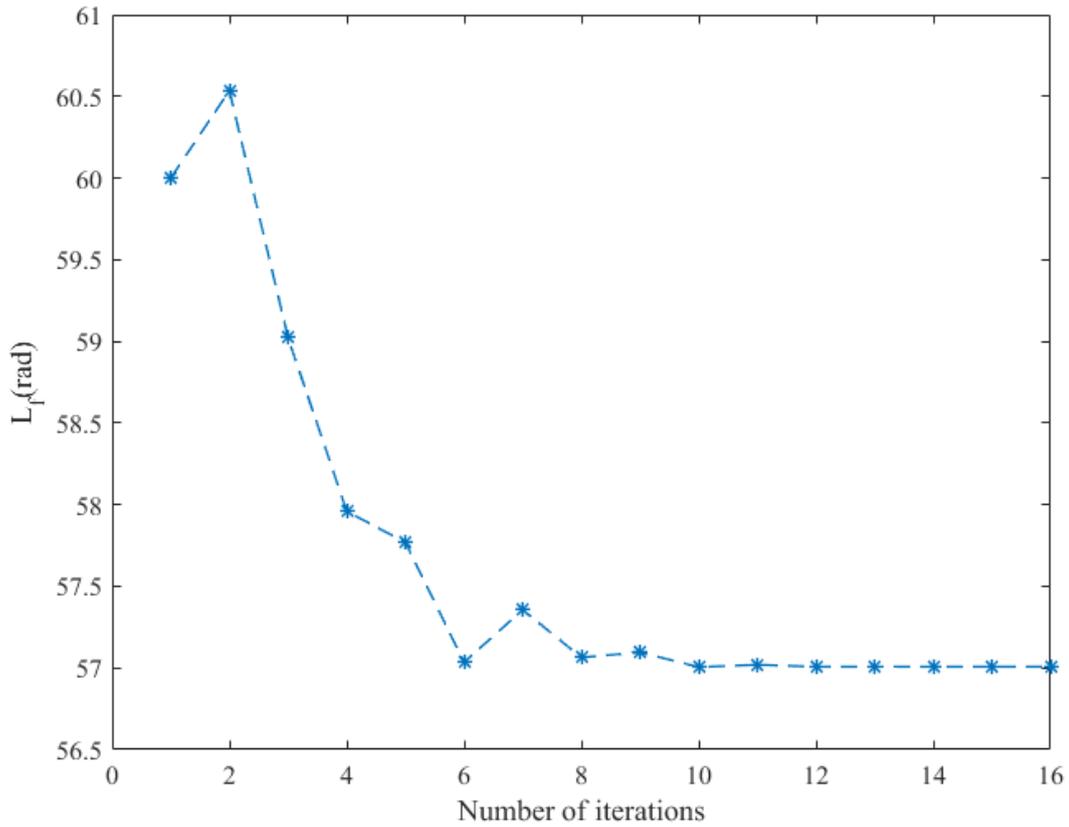

**Fig. 8** Evolution of the final true longitude during the optimization process for the symplectic method

**4.4 Low Earth Orbit Spacecraft Rendezvous**

Consider a spacecraft rendezvous mission in Low Earth Orbit (LEO): the chaser satellite starts from a sun-synchronous orbit and transfer to another sun-synchronous orbit to rendezvous with the target satellite. Unlike the last two numerical examples, the spacecraft dynamics around Earth include the J2 perturbation, which makes the optimal control problem much challenging to solve (Zhao et al. 2017). This example illustrates that, the symplectic method can also be applied to optimize low-thrust trajectories with a very large number of revolutions within perturbed dynamics. The specific impulse $I_{sp}$ fixed to 3800 s and the initial mass of the chaser spacecraft is equal to 1500 kg. The maximum thrust magnitude is 0.33N.

The initial state vector of the chaser spacecraft is specified in terms of equinoctial orbit elements as:

$$p_0(t_0) = 6899.4468 \text{ km}, \quad f_0(t_0) = -0.00008344, \quad g_0(t_0) = 0.00067183$$
$$h_0(t_0) = -0.06749657, \quad k_0(t_0) = -1.13743783, \quad L_0(t_0) = 1.85174464 \text{ Rad}$$
(76)

The initial state of the target spacecraft is also specified in terms of equinoctial orbit elements as:

$$p_1(t_0) = 6897.4283 \text{ km}, \quad f_1(t_0) = -0.00026998, \quad g_1(t_0) = 0.00040531$$
$$h_1(t_0) = -0.06750251, \quad k_1(t_0) = -1.13753794, \quad L_1(t_0) = 1.99497980 \text{ Rad}$$
(77)

It should be noted that, the target spacecraft is subject only to the Earth's gravity, while the chaser spacecraft is subject both Earth's gravity and the thrust of its own electric propulsion system.

The characteristics quantities that normalize the problem are changed to reflect the fact that the Earth is, now, the central body (in contrast to the previous examples). The characteristic length is set to $L = 6878.137$ km. Then, the characteristic time can be defined as $T = (L^3 / \mu_e)^{0.5} = 903.52$ s, so to make the normalized $\mu_e$ equal to 1. The initial spacecraft mass is chosen as the characteristic mass.

Initially, we set the transfer time of flight to 2 days, which corresponds to a trajectory with 30 revolutions. Again, we solve this numerical example with different number of trajectory arcs. The converged results are listed in Table 7. Observations from section 4.2 are still valid in Table 7. That is, the symplectic method may reach relatively high accuracy with 3 or 4 arcs, and the accuracy of the symplectic method can further improve when more arcs are added. As a reference, the optimal solution for 100 intervals is portrayed through Figures. 9, 10 and 11.

**Table 7** Converged solutions for 2-day LEO spacecraft rendezvous with different number of arcs

| Number of Arcs | Number of Iterations | Computational Time(s) | Final Mass(kg) |
|---|---|---|---|
| 45 | 4 | 1.808894 | 1499.553504 |
| 50 | 3 | 2.172762 | 1499.764308 |
| 60 | 3 | 3.512631 | 1499.807556 |
| 70 | 3 | 4.891646 | 1499.760483 |
| 80 | 3 | 6.768178 | 1499.761093 |
| 90 | 3 | 8.838494 | 1499.761339 |
| 100 | 3 | 11.512130 | 1499.761217 |
| 200 | 3 | 70.278182 | 1499.761266 |
| 300 | 3 | 216.140373 | 1499.761265 |

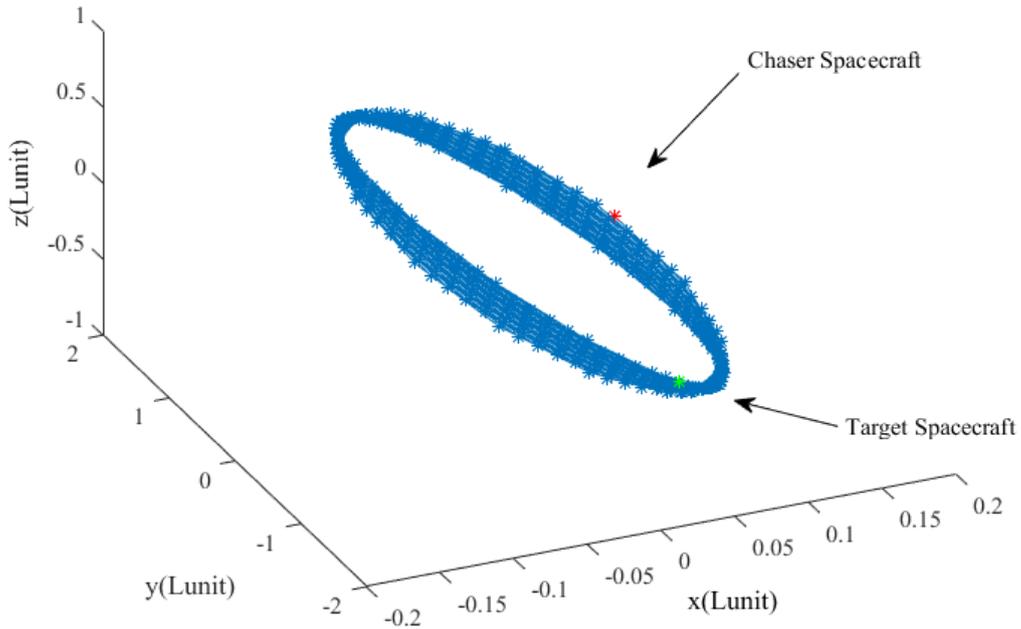

**Fig. 9** Low-thrust trajectory with 30 revolutions for the chaser spacecraft

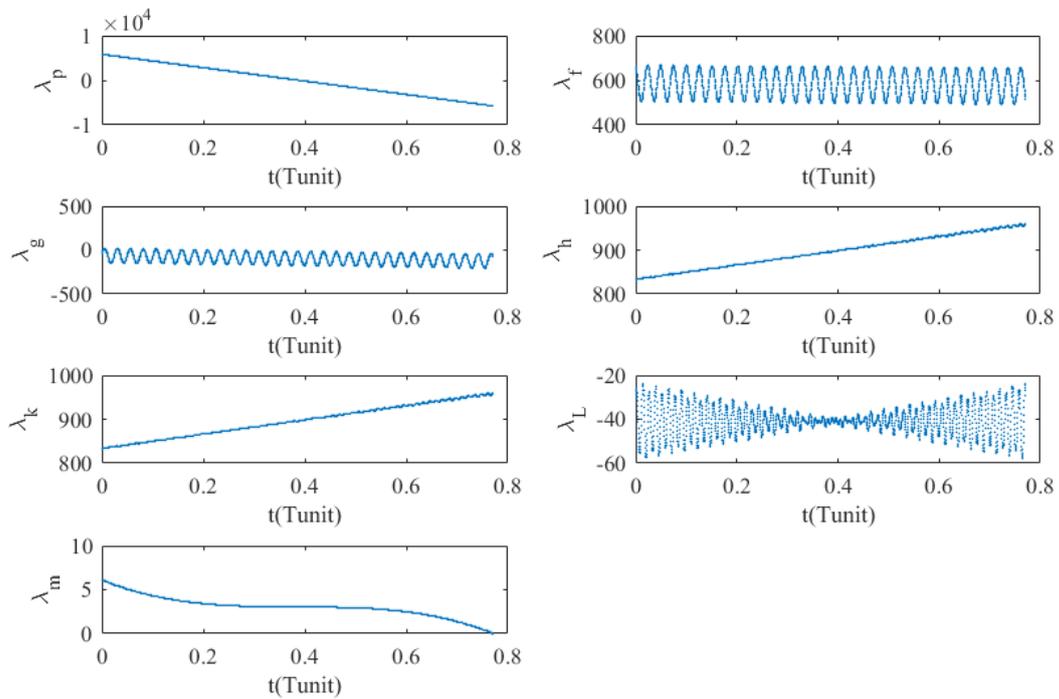

**Fig. 10** Costate variables time histories of the chaser spacecraft along a 30 revolutions trajectory

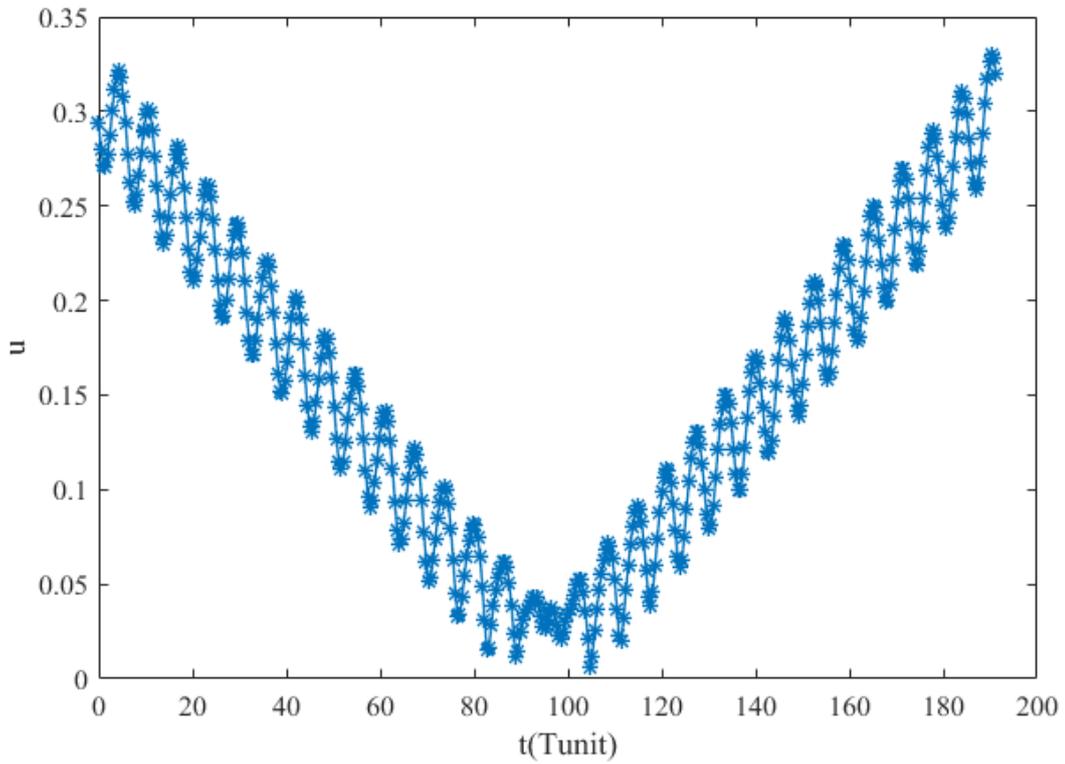

**Fig. 11** Optimal thrust profile of the chaser spacecraft along a 30 revolutions trajectory

Next, we consider a longer time of flight, i.e., 15 days which corresponds to a 228-revolutions trajectory. Solving low-thrust trajectory with such a large number of revolutions is considered a challenging problem. The optimal solution can be successfully obtained by the symplectic method, when a good initial guess is supplied. The converged optimal solutions are listed in Table 8. The CPU cost could be further reduced by improving code quality. For completeness, the trajectory, the costate variables and the thrust profile are depicted in Figures 12, 13 and 14. This example supplies preliminary evidence that the proposed symplectic method is a promising tool to optimize low-thrust transfers with a large number of revolutions.

**Table 7** Converged solutions for 15-day LEO spacecraft rendezvous with different number of arcs

| Number of Arcs | Number of Iterations | Computational Time(s) | Final Mass(kg) |
| --- | --- | --- | --- |
| 600 | 3 | 1719.7184 | 1499.943624 |
| 800 | 3 | 3476.3645 | 1499.964308 |

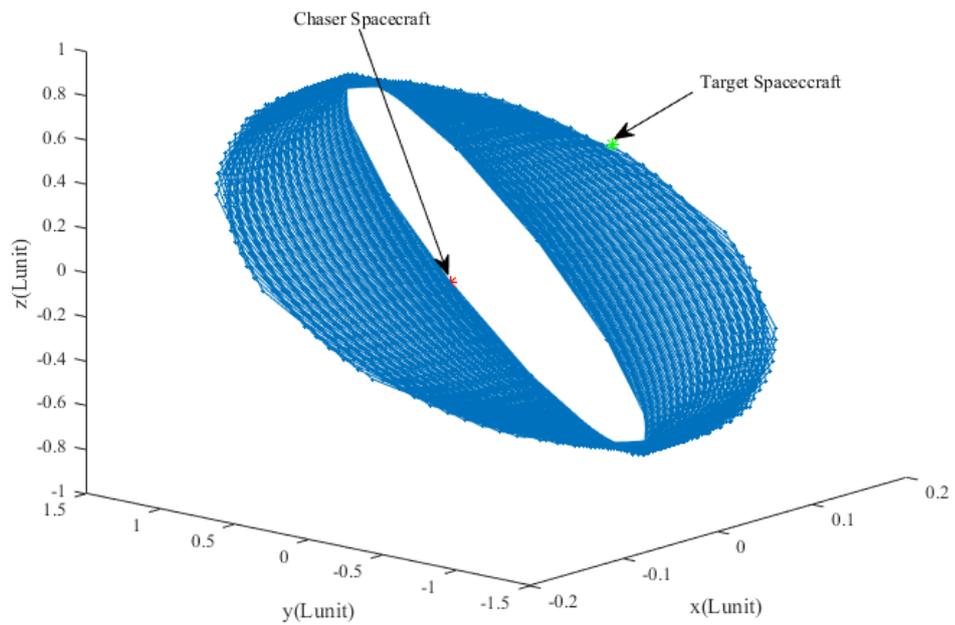

**Fig. 12** Low-thrust trajectory of the chaser spacecraft with 228 revolutions

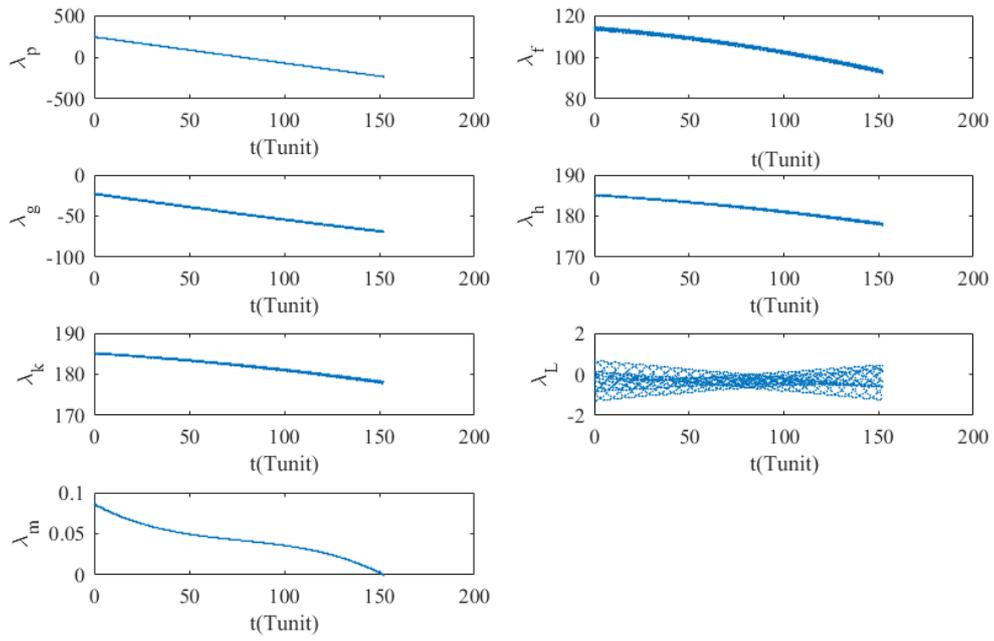

**Fig. 13** Costate variables time histories of the chaser spacecraft along a

228 revolutions trajectory

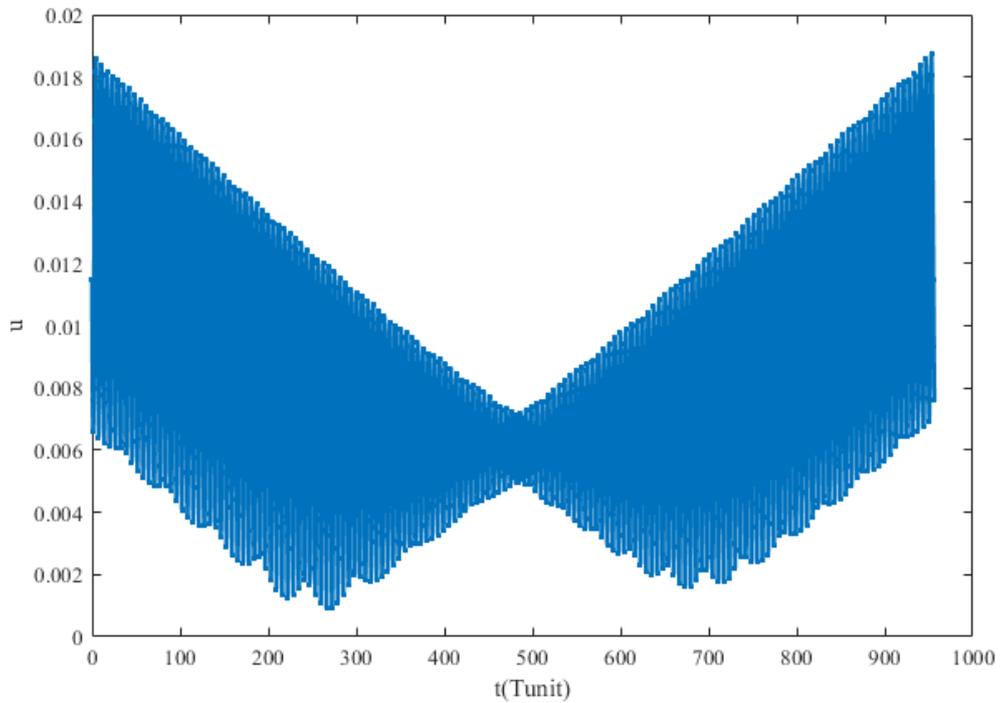

**Fig. 14** Optimal thrust profile of the chaser spacecraft long a 228 revolutions trajectory

## 5. Conclusion

A symplectic method is presented in this paper to optimize multi-revolution low-thrust orbit transfer trajectories. To reduce the oscillatory nature of the Cartesian coordinates along spiraling trajectories with multiple revolutions, osculating equinoctial elements are chosen to describe the motion of the spacecraft. In addition, an initial reference solution is given to accelerate the optimization process. These two techniques may enable the symplectic method to converge rapidly, when it is applied to the optimization of orbit transfer with multiple revolutions.

A representative renhdezvous problem from the Earth to Venus is successfully solved by the proposed method. The accuracy and optimality of the symplectic method are demonstrated by comparison with known results obtained from an indirect method. In addition, the relationship between the number of intervals and the accuracy attainable with the symplectic method is discussed, and may be a reference for future research. The symplectic method is also compared to the well-known SNPOT solver. In optimizing an orbit transfer between two circular orbits, which serves as a benchmark problem, the symplectic method is significantly faster than SNOPT in terms of computational time. In addition, compared to SNOPT, the proposed method can produce a reasonable approximation of the continuous solution with fewer discretization points. Finally, low Earth orbit spacecraft rendezvous with a very large number of revolutions are successfully solved by the proposed symplectic method, within J2-perurbed orbit dynamics. In conclusion, the symplectic methods prove valid in solving known problems and seem to behave well when applied to more complex dynamics. In future work, we envision the application of symplectic methods to optimize more complex transfers within higher fidelity environments.

**Acknowledgment**: This work is supported by the National Natural Science Foundation of China (No. 11672146 and No. 11432001) and the 2015 Chinese National Postdoctoral International Exchange Program.